\documentclass[sigconf]{acmart}
\AtBeginDocument{%
  }

\copyrightyear{2026}
\acmYear{2026}
\setcopyright{cc}
\setcctype{by}
\acmConference[KDD '26]{Proceedings of the 32nd ACM SIGKDD Conference on Knowledge Discovery and Data Mining V.2}{August 09--13, 2026}{Jeju Island, Republic of Korea}
\acmBooktitle{Proceedings of the 32nd ACM SIGKDD Conference on Knowledge Discovery and Data Mining V.2 (KDD '26), August 09--13, 2026, Jeju Island, Republic of Korea}
\acmDOI{10.1145/3770855.3818037}
\acmISBN{979-8-4007-2259-2/2026/08}

\usepackage{xcolor}
\usepackage{colortbl}
\usepackage{enumitem}

\definecolor{purplebase}{RGB}{160, 60, 200}  

\newlength{\papercontriblabelwidth}
\newcommand{\papercontribution}[2]{%
  \par\noindent
  \makebox[\papercontriblabelwidth][l]{\textbf{#1:}}%
  \parbox[t]{\dimexpr\linewidth-\papercontriblabelwidth\relax}{#2}%
  \par
}
\begin{document}

\title{Rewarding Engagement and Personalization in Popularity-Based Rankings Amplifies Extremism and Polarization}



\author{Jacopo D'Ignazi}
\orcid{0000-0003-2843-5279}
\email{jacopo.dignazi@upf.edu}
\affiliation{%
  \institution{Universitat Pompeu Fabra}
  \city{Barcelona}
  \country{Spain}
}

\author{Emma Fraxanet}
\orcid{0000-0002-1647-7300}
\email{emmafraxanet@gmail.com}
\affiliation{%
  \institution{Barcelona Supercomputing Center}
  \city{Barcelona}
  \country{Spain}
}

\author{Andreas Kaltenbrunner}
\orcid{0000-0002-2271-3066}
\email{andreas.kaltenbrunner@upf.edu}
\affiliation{%
  \institution{Universitat Pompeu Fabra}
  \city{Barcelona}
  \country{Spain}
}

\author{Gaël Le Mens}
\orcid{0000-0003-4800-0598}
\email{gael.le-mens@upf.edu}
\affiliation{%
  \institution{Universitat Pompeu Fabra, Barcelona School of Economics and UPF-BSM}
  \city{Barcelona}
  \country{Spain}
}

\author{Fabrizio Germano}
\orcid{0000-0002-6211-4519}
\email{fabrizio.germano@upf.edu}
\affiliation{%
  \institution{Universitat Pompeu Fabra and Barcelona School of Economics}
  \city{Barcelona}
  \country{Spain}
}

\author{Vicenç Gómez}
\orcid{0000-0001-5146-7645}
\email{vicen.gomez@upf.edu}
\affiliation{%
  \institution{Universitat Pompeu Fabra}
  \city{Barcelona}
  \country{Spain}
}

\renewcommand{\shortauthors}{Jacopo D’Ignazi et al.}

\begin{abstract}
Despite extensive research, the mechanisms through which online platforms shape extremism and polarization remain poorly understood, largely because observational settings make it difficult to identify causal effects. We propose a framework that analyses a specific mechanism: the feedback loop between popularity-based rankings and user behavior (e.g., partisan sorting and attention dynamics), and its consequences for content prominence.
Our framework formalizes this feedback loop using a dynamical, popularity-based ranking model with a small set of interpretable parameters, and evaluates it through simulations and interactive experiments with hundreds of human participants. When trained on real-world data, the model reproduces well-established interaction patterns with ranked content: (1) users exhibit position bias, (2) prefer like-minded content, and (3) more extreme users engage more actively. Using this framework, we show that when platforms reward \emph{active} engagement and implement \emph{personalized} rankings, users are systematically driven toward increasingly extremist and polarized news consumption across a wide range of parameter settings.

\end{abstract}




\maketitle

\newcommand\kddavailabilityurl{https://doi.org/10.5281/zenodo.20475404}
\ifdefempty{\kddavailabilityurl}{}{%
\begingroup\small\noindent\raggedright\textbf{Resource Availability:}\\
Code and processed data needed to reproduce experiments, simulations, analyses, and figures are available at \url{\kddavailabilityurl}.
\endgroup
}

\section{Introduction}

Online platforms are central to political communication and public discourse. A vast literature has examined their potential in shaping opinion and information dissemination \cite{zhuravskaya2020political}, fueling extremism~\cite{finkel2020political}, and generating ideologically segregated spaces known as echo chambers \cite{bruns2019filter, cinelli2021echo}. 

The mechanisms that drive these effects, however, remain contested \cite{haidt_bail_social_media_dysfunction}. Some studies report limited attitudinal effects of altering exposure during large platform interventions~\cite{guess2023social,nyhan2023like}, while other work shows that exposure to opposing views can backfire and increase partisan hostility \cite{bail2018exposure}. Complementary evidence on Facebook indicates that individual choices, more than algorithmic ranking, reduce cross-cutting exposure, and that position in the ranked list strongly shapes clicking behavior~\cite{bakshy2015exposure}. However, when different levels of content filtering are examined, it is shown that ideological segregation increases in all steps, from algorithmic curation to user engagement~\cite{gonzalez2023asymmetric}. Similarly, field studies around the 2020 U.S. election reveal that algorithmic choices strongly shape exposure and engagement, yet the short-run effects on attitudes are limited \cite{gonzalez2023asymmetric,guess2023social,guess2023reshares,nyhan2023like}. These mixed findings underscore the importance of understanding the interaction between user behavior and algorithmic design choices~\cite{lorenz2023systematic}.

Over the last few years, a body of empirical research has documented regularities in user interaction with news and ranked content. Examples include position bias, where items ranked higher attract disproportionately more clicks~\cite{bakshy2015exposure,joachims05,salganik06,epstein15,le2018endogenous,germano2019few}, confirmatory tendencies, where individuals prefer content aligned with their prior beliefs~\cite{bakshy2015exposure, fraxanet-EPJ25,flaxman2016filter,bail2018exposure,konovalova2023social}, and a U-shaped distribution of activity, where users with extreme views engage more intensively through liking, sharing, or commenting \cite{bakshy2015exposure,fraxanet-EPJ25,lee2025network}.
Regarding algorithmic design, user engagement (measured through clicks, shares, or reactions) is a key criterion for optimizing ranking and recommendation systems~\cite{guess2023social,guess2023reshares,nyhan2023like,huszar2022algorithmic}, often at the expense of exposing users to more polarizing or toxic content~\cite{chavalarias2024can}. In practice, engagement rewards and personalization are usually implemented together rather than as isolated mechanisms. For example, Facebook's Meaningful Social Interactions update can be interpreted as simultaneously increasing the weight of active engagement and the degree of personalization in the ranking process~\cite{meta_2018,wsj_2021}.

\begin{figure}[!t]
     \centering
     \includegraphics[width=\columnwidth]{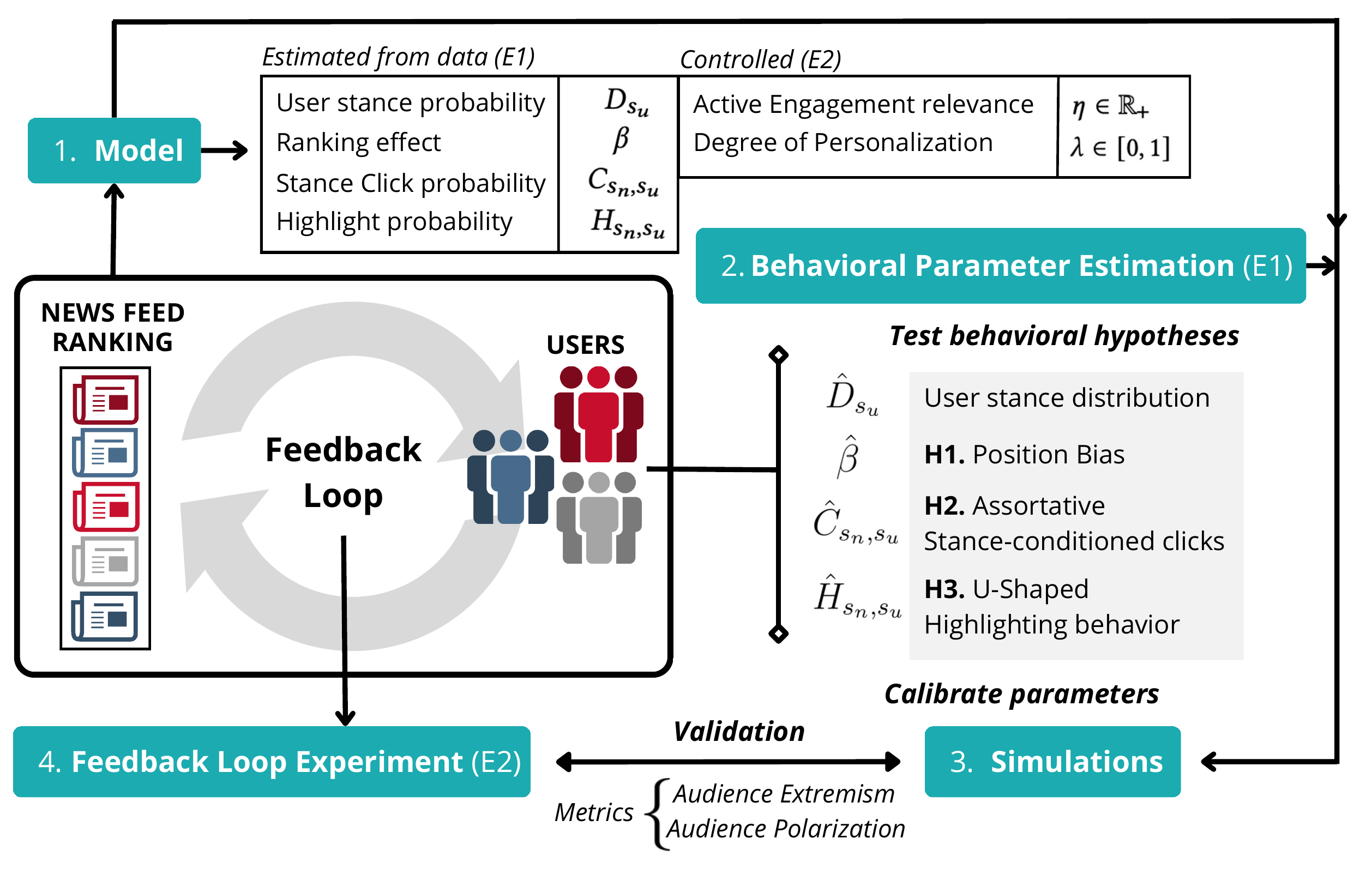}
     \caption{Overview of the modeling and experimental framework to study the feedback loop between engagement-based rankings and user behavior. Its four main components are: (i) the model parameters, (ii) the behavioral parameter estimation (E1) and corresponding quantities, (iii) the data-driven simulation step, and (iv) the feedback loop experiment (E2).}
     \label{fig:overview}
 \end{figure}

Building on this evidence, we propose a methodological framework (summarized in Fig.~\ref{fig:overview}) that specifically targets the mechanism behind the feedback loop between user behavior and ranking algorithms: 
(1) We propose a simple model that formalizes the aforementioned mechanism, (2) calibrate its parameters using behavioral data, (3) analyze its predictions via simulations, and (4) test them in a human-in-the-loop experiment. Our framework thus provides three main contributions: 

\papercontribution{C1}{A data-driven approach for simulating a popularity-based ranking system, validated against experiments with human participants.}

\papercontribution{C2}{A controlled environment for experiments with human participants to examine the causal effect of an engagement-driven ranking regime on user consumption.}

\papercontribution{C3}{The identification of a mechanism driving polarized and ideologically extreme consumption, rooted in the interaction between user behavior and algorithmic ranking.}


In our experiment, a ranking system displays personalized lists of news to users, 
who interact through clicks and highlights. These interactions update item popularity, reshaping the ranking shown to subsequent users and creating a feedback loop between user behavior and algorithmic visibility. 
In essence, we find that when a ranking algorithm gives more weight to engagement and personalization, content consumed by users with more extreme views becomes more popular. This content then becomes more visible to less extreme users, who consequently consume it more often. The outcome of this feedback loop is thus the amplification of both audience polarization and ideologically extreme consumption~\cite{mangold2022metrics}. 

Our study is focused on the effect of the feedback loop on \emph{consumption} patterns, rather than attitudinal change. While this work does not assess downstream effects such as durable opinion change, increased affective polarization or ideological radicalization, prior work suggests that selective exposure and engagement-based amplification can reinforce existing views and shape public discourse over time~\citep{doi:10.1126/science.adu5584, bail2018exposure,flaxman2016filter,lorenz2023systematic}.

\section{Related Work}

Our work relates to three different lines of research:

\emph{Evidence from Online Platforms and Field Experiments.} —
A growing empirical literature examines how platform design and ranking choices shape exposure, engagement, and attitudes across major platforms~\cite{whittaker2021recommender}. Randomized controlled experiments conducted during the 2020 U.S. presidential election on Facebook and Instagram demonstrate that feed-ranking and reshare features strongly shaped user exposure to and engagement with political content \cite{gonzalez2023asymmetric,guess2023social,guess2023reshares,nyhan2023like,allcott2024effects}. Beyond Facebook, evidence from Twitter and Google Search shows that algorithmic design also influences amplification and partisan engagement \cite{huszar2022algorithmic,robertson2023users}, underscoring the broader role of ranking choices in shaping online consumption. These experiments also sparked debate about context and validity~\cite{bagchi2024social,guess2023social}. A recent study proposes a method to test feed-ranking interventions without platform collaboration, using a browser extension~\cite{doi:10.1126/science.adu5584}. They show that experimentally up- or down-ranking posts expressing anti-democratic attitudes and partisan animosity, respectively, increases or decreases affective polarization. In contrast, our approach focuses on how engagement-driven ranking dynamics reshape audience polarization over time.

Our work complements this literature by formalizing a simple mechanism through which engagement-driven personalized ranking can amplify audience polarization and ideologically extreme consumption. We test this mechanism in controlled human-in-the-loop experiments with dynamically updating rankings, establishing the causal effect of a combined personalization-and-engagement regime relative to a neutral baseline.

\emph{Mechanisms and models (theory and simulation).}—
Several analytical and simulation-based models have been proposed to study how user–algorithm feedback drives polarization.  
Analytical approaches~\cite{acemoglu2024model} formalize opinion and network dynamics under algorithmic influence, while agent-based and simulation studies~\cite{tornberg2023simulating,chavalarias2023can,hsu2025game} model adaptive interactions between users and recommender algorithms, including engagement optimization and misinformation spread. Work on feedback and personalization mechanisms~\cite{mansoury2020feedback,santos2021link,liu2021interaction,piao2023human} further shows how user–algorithm loops amplify popularity bias, polarization, and information cocoons.  

The model proposed here is most similar to those recently introduced in~\cite{SOBBRIO201443,germano2019few,germano2020opinion,germano2025ranking}. In particular, we build on~\cite{germano2025ranking}, which provides the theoretical mechanism motivating our study. Our contribution is to adapt this mechanism to a data-driven and experimentally testable setting, using discrete stance categories, explicitly including center users, and estimating behavioral components from human data. The behavioral assumptions are tested in a first controlled experiment, while a second controlled experiment provides causal evidence on the joint effect of personalization and engagement reward on audience polarization and ideologically extreme consumption, empirically corroborating the predictions of~\cite{germano2025ranking}. This focus on ranking dynamics complements work on polarization and extremization that studies how users adapt their behavior over time to feedback from other users~\citep{konovalova2023opinion,scholl2024}.

\emph{Algorithmic Design for Mitigating Adverse Effects.} —
Developing recommendation and ranking algorithms that mitigate adverse societal effects has emerged as a central research challenge. Examples include consequential systems that balance immediate utility and long-term welfare~\citep{tabibian2020design}, the incorporation of partisan audience diversity as a reliability signal in collaborative filtering~\citep{bhadani2022political, wojcik2022birdwatch},  attention mechanisms that emphasize topic-specific terms while down-weighting polarized language~\citep{shivaram2022reducing}, and the provision of negative quantitative feedback via a ``dislike'' button~\citep{konovalova2023social}. More recent approaches include online feedback control to trade off click-through rates with harm~\citep{CheeKDD24}, the use of conformal risk control to provably bound unwanted content in personalized recommendations~\cite{cr_control}, graph-based strategies to disrupt radicalization pathways~\citep{fabbri2022rewiring}, and model-side techniques to limit attention polarization \citep{liu2024deanchor}.

Whereas these frameworks emphasize algorithmic performance, our model offers a principled understanding of how user behavior and ranking mechanisms coevolve, which can guide the design of more transparent and theoretically grounded interventions.

\section{Methodological Framework}

Our framework consists of four components, schematized in Fig.~\ref{fig:overview}. First, we introduce a {\bf model} that formalizes user interaction with a ranked news list, and ranking updates based on engagement.

Second, we designed and implemented a simple online platform to conduct a first controlled experiment with human participants; we call this experimental setup {\bf Behavioral parameter estimation~(E1)}, as it is aimed at verifying behavioral hypotheses and informs the data-driven simulations.

Third, we use the behavioral estimates and the model to {\bf simulate} scenarios with different ranking algorithms, and analyze how consumption patterns are affected.

Finally, we use two of these parameter settings to carry out a second controlled experiment. We call this stage {\bf feedback loop experiment (E2)}. Motivated by the platform setting discussed above, the experiment compares a neutral baseline, with no personalization and no reward for active engagement, to an engagement-driven regime that combines strong personalization with a strong reward for active engagement. Thus, E2 tests the joint effect of these two algorithmic choices, while their separate roles are examined through simulations that vary $\lambda$ and $\eta$ systematically. We then analyze the outcome of the two scenarios, perform statistical tests to assess their differences, and compare the results with simulated predictions.

\subsection{Model}
\label{sec:model}

Our model consists of a {\bf user behavior model}, capturing how individuals select and engage with content, and a {\bf ranking algorithm}, which updates the ranking based on the 
user interactions.

Both news items and users are assigned a political stance on a five-point spectrum representing ideological alignment: extreme left, moderate left, center, moderate right, and extreme right. We define the stance set as 
\begin{equation}\label{eq:stance_set}
     S=\{-2,-1,0,1,2\},
\end{equation}
where negative values correspond to left-leaning positions.

Each ranked list contains $N$ news items, each assigned a stance
\begin{equation}
    s_n \in S \quad \mbox{with} \quad n \in \{1,\dots,N\}.
       \nonumber
\end{equation}
Similarly, we consider $U$ users, assumed to arrive sequentially to interact with the ranking, each assigned a stance
\begin{equation}
    s_u \in S \quad \mbox{with} \quad u \in \{1,\dots,U\}.
       \nonumber
\end{equation}

We assume that $s_n$ and $s_u$ are drawn from distributions $D_{s_n}$ and~$D_{s_u}$, respectively. The ranked list is initialized with a uniform representation of news stances $D_{s_n}$, while the user stance distribution $D_{s_u}$ is empirically estimated from experimental data.

\subsubsection{User behavior}\label{sec:user_behavior}

At each time step, the platform displays a ranked list of news items, denoted by $\overline{r}$, which specifies the permutation of item indices that defines their order in the ranking. The position (rank) of a specific item $n$ in this list is denoted by $r_n$.

Each user interacts with this list by first clicking on one
of the news items and then deciding whether to actively engage with it (e.g., by liking or sharing). We denote this optional form of active engagement \emph{highlighting}. We model the click probability as the result of two independent factors:

\textbf{Position bias}, which captures the tendency of higher-ranked items to attract more attention, consistent with existing evidence on clicking behavior~\citep {bakshy2015exposure,joachims05,salganik06,epstein15,le2018endogenous,germano2019few} ({\bf Hypothesis~H1}):
\begin{equation}
\label{eq:beta}
    R(r_n):=\beta^{N-r_n}, \qquad  \beta\geq 1,
\end{equation}
so that an item at rank $r_n$ is $\beta$ times more likely to be clicked than an item at rank $r_{n+1}$, all else being equal. 
The exponential form gives~$\beta$ a direct interpretation as the multiplicative change in attention when item $n$ moves down one position, since $R(r_n)/R(r_n+1)=\beta$, following prior work~\cite{germano2019few}. We also repeated the analysis with a power-law decay, as used in feed recommendation models~\cite{jeunen2023probabilistic}, and obtained the same conclusions.

\textbf{Stance-conditioned baseline click probabilities.}
For each user stance $s_u \in S$, the clicked item’s stance is modeled as a categorical variable over $S$ with parameter vector $C_{\cdot,s_u}$:
 \begin{equation}\label{eq:C} 
C_{s_n,s_u} := P(\text{click on stance } s_n \mid s_u), \qquad \sum_{s_n \in S} C_{s_n,s_u} = 1.
\end{equation}


We conjecture that when the news item and the user have the same stance $s_n=s_u$, $C_{s_n,s_u}$ is larger than chance ($1/5$), consistent with the finding that users prefer content aligned with their prior beliefs and attitudes~\citep{bakshy2015exposure,white2015belief,flaxman2016filter,bail2018exposure,konovalova2023social} ({\bf Hypothesis~H2}).

Combining both factors, the probability that a user with stance~$s_u$ clicks on item $n$ when presented with ranking $\overline{r}$ is
\begin{equation}\label{eq:ranking_prob}
    P(n \text{ clicked} \mid \overline{r}, s_u)
    := \frac{R(r_n)\,C_{s_n,s_u}}{\sum_{n'\in\{1,\hdots,N\}} 
    R(r_{n'})\,C_{s_{n'},s_u}}.
\end{equation}

\textbf{Highlighting behavior.}
After a click, a user may optionally highlight the item. For stances $(s_n,s_u)$, we define
\[
H_{s_n,s_u} := P(\text{highlight}\mid \text{click}, s_n, s_u),
\]
so the post-click highlight is Bernoulli with parameter $H_{s_n,s_u}$.


We conjecture that the overall probability of active engagement varies with user stance in a U-shaped pattern, with extreme users more likely to highlight content than moderates, consistent with existing evidence~\cite{bakshy2015exposure,fraxanet-EPJ25,lee2025network} (\textbf{Hypothesis~H3}).

\subsubsection{Ranking algorithm}\label{sec:ranking_algorithm}

We introduce the update equations defining how the ranking evolves from user interactions. The model assumes a popularity-based ranking, where more popular items appear higher. Item popularity depends on two factors:
\begin{itemize}
    \item \emph{Active engagement relevance} — highlighted news receive an additional reward in popularity, controlled by parameter $\eta$.
    \item \emph{Personalization degree} — the mixing between ranking seen by left, center, and right users, controlled by parameter $\lambda$.
\end{itemize}

We implement a basic form of personalization, defined at the level of user groups according to political stance:
\begin{equation}
    L=\{-2,-1\}, \qquad C=\{0\}, \qquad R=\{1,2\}, 
    \qquad g \in \{L,C,R\}.
       \nonumber
\end{equation}

For each user group $g$, we define the popularity of news item $n$ at time $t$ as 
\begin{equation}
    p^{g}_{n}(t) \in \mathbb{R}_+, 
    \nonumber
\end{equation}
so that more popular items are displayed in higher positions within the ranking.
We initialize $p^{g}_{n}(0)=0$ for all $n$, using a random initial ordering to break ties.

Let $u$ denote the user arriving at the platform at time-step~$t$, and let $n$ be the clicked news item. The popularity increment of the clicked item $n$ in each group $g\in \{L,C,R\}$ is defined as follows:
\begin{equation}\label{eq:delta_pop}
    \Delta p^g_{n}(t) :=
    \begin{cases}
        1 & \text{if } s_u\in g \text{ and } n \text{ not highlighted},\\
        1+\eta & \text{if } s_u\in g \text{ and } n \text{ highlighted},\\
        (1-\lambda) & \text{if } s_u\notin g \text{ and } n \text{ not highlighted},\\
        (1-\lambda)(1+\eta) & \text{if } s_u\notin g \text{ and } n \text{ highlighted}.
    \end{cases}
\end{equation}
Non-clicked items receive no popularity increment.

Note that for $\eta=0$, the popularity depends only on clicks, and active engagement (highlighting) is not promoted on the platform. Similarly, for $\lambda=0$, there is no personalization, and all users are exposed to the same ranking. In contrast, for $\lambda=1$, group-wise personalization is maximal, and readers are exposed to a ranking that ignores the activity of users with a different stance. This parametrization explores four scenarios, corresponding to extreme cases:
\begin{enumerate}
    \item $\lambda=0, \eta=0$: \textbf{no} personalization, \textbf{no} highlight reward
    \item $\lambda=0, \eta=100$: \textbf{no} personalization, \textbf{full} highlight reward
    \item $\lambda=1, \eta=0$: \textbf{full} personalization, \textbf{no} highlight reward
    \item $\lambda=1, \eta=100$: \textbf{full} personalization, \textbf{full} highlight reward
\end{enumerate}


Fig.~\ref{fig:overview} summarizes the model parameters: user behavior is characterized by $D_{s_u}$, $\beta$, $C_{s_n,s_u}$, and $H_{s_n,s_u}$, which are estimated from data (Sections~\ref{sec:E1} and~\ref{sec:res_E1}), while the platform parameters $\eta$ and $\lambda$ are experimentally controlled (Sections~\ref{sec:E2} and~\ref{sec:res_E2}).

\subsubsection{Metrics}\label{sec:metrics}

\begin{figure}[t!]
    \centering
    \includegraphics[width=\columnwidth]{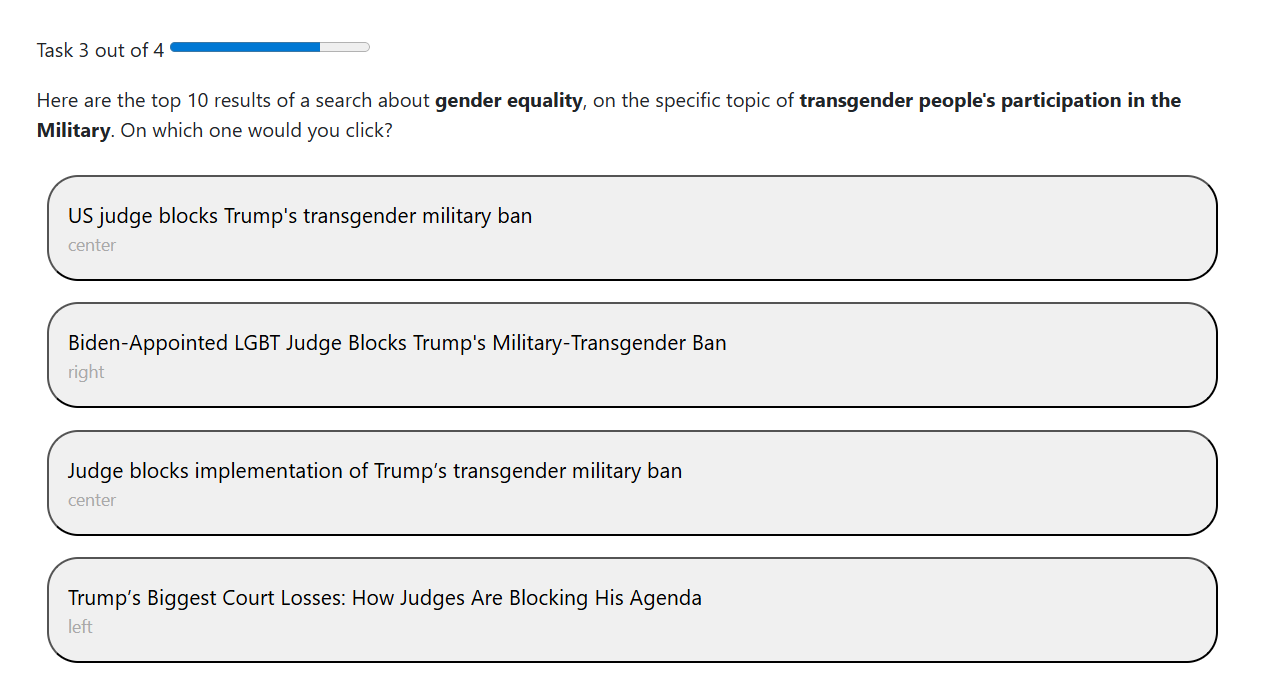}
    \caption{Example page in the "Ranked List and Clicking" step. Participants can scroll through up to 10 articles and were neither required nor encouraged to inspect all items; the instruction "On which one would you click?" permits natural scrolling.}
    \label{appendix:fig:ss_2}
\end{figure}
We focus our analysis on the \emph{extremism} and \emph{polarization} of the system induced by the ranking of news items clicked at time $t$. These metrics are computed based on the clicks received over a time window $W$ of size $w$,
 where $\tau \in W$ if $\tau \in (t-w,t]$. Let $\hat{s}_n(\tau)$ denote the stance of the clicked news in any partition at time~$\tau$. We define {\bf Audience Extremism} as
\begin{equation}\label{eq:ext}
Ext(t)=\langle\vert\hat{s}_n(\tau)\vert\rangle _{\tau\in W}
\end{equation}
which corresponds to the average distance of the clicked items' stance from the center.

For polarization, we instead measure the difference between the average stance of the news items clicked by users in the right and left partitions. In this case, $W=W_L \cup W_C \cup W_R$, so the averages on each partition are calculated on a subset of the time window $W$ corresponding to clicks from users in that partition. We thus define {\bf Audience Polarization}~\cite{mangold2022metrics} as
\begin{equation}\label{eq:pol}
Pol(t)=\langle\hat{s}_n(\tau)\rangle_{\tau\in W_R} - \langle \hat{s}_n(\tau) \rangle_{\tau\in W_L}.
\end{equation}

By construction, $Pol(t)$ focuses on the separation between left- and right-leaning audiences, which is the main axis of ideological polarization in our setting. Center users are not used to define this metric, but they contribute to $Ext(t)$ and to the full distribution of clicked stances analyzed below. Including center users through pairwise distances between group-level average clicked stances leads to the same qualitative ordering of conditions.

We focus on audience extremism and polarization after the ranking dynamics have empirically stabilized. Extended simulations show that the metrics stabilize after an initial transient period, supporting the use of the final interaction window in the human experiment. For the time evolution of these metrics, see Sec.~\ref{sec:results_simulation}.


\subsection{Simulations}\label{sec:simulation}
Given the model parameter estimates from our first experiment (see later in Sec.~\ref{sec:E1}) \, we simulated the model to predict Audience Extremism and Audience Polarization (Sec.~\ref{sec:metrics}), when the ranking is dynamically updated as described in Sec.~\ref{sec:ranking_algorithm}.
We first explored the parameter space in small increments of $\eta\in[0,100]$ and $\lambda \in [0,1]$, using $1{,}000$ independent simulations for each algorithmic configuration. Each independent simulation was initialized with a random ranked list. We then focused on the four extreme cases of Sec.~\ref{sec:ranking_algorithm}, using estimates obtained for each individual topic.

Each independent simulation consisted of the sequential interaction of $250$ synthetic users with the ranked list of news items. Their stance was sampled from $\hat{D}$, while their behavior was determined by the topic-specific estimates $\hat{\beta}$, $\hat{C}$, and $\hat{H}$ obtained from E1.

We used $1{,}000$ independent simulations to obtain stable averages, while $250$ synthetic users were sufficient for the metrics to stabilize (see Sec.~\ref{sec:results_simulation}). This also matches the approximate number of participant interactions per trajectory in E2 (see Sec.~\ref{sec:E2}).

\begin{figure*}[ht]
    \centering
    \includegraphics[width=.24\textwidth]{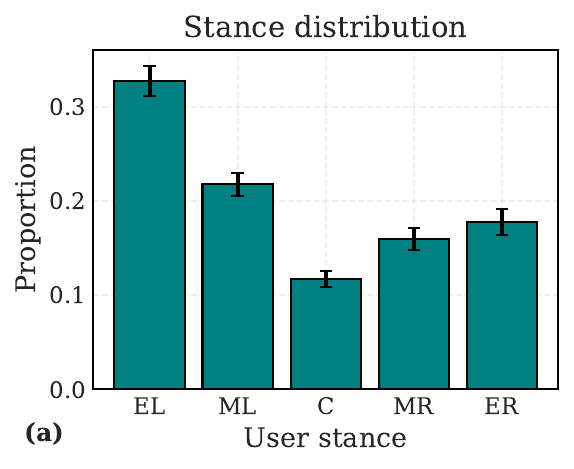}
    \hspace{2pt}
    \includegraphics[width=.24\textwidth]{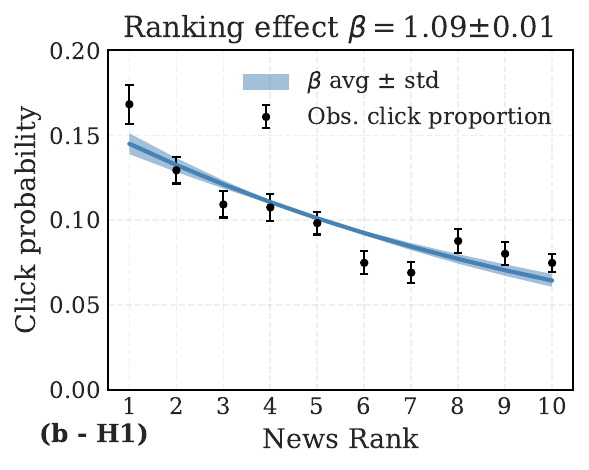}
    \hspace{2pt}
    \includegraphics[width=.24\textwidth]{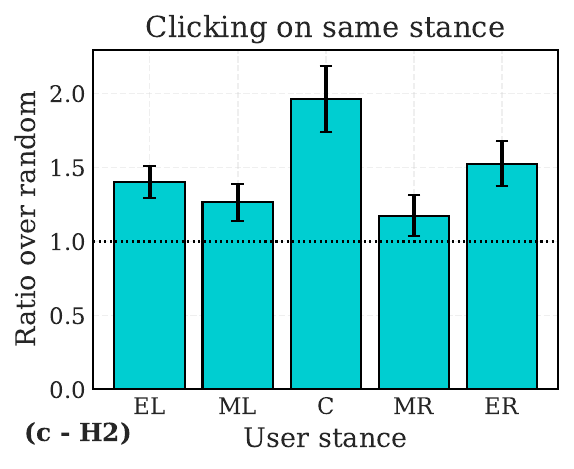}
    \hspace{2pt}
    \includegraphics[width=.24\textwidth]{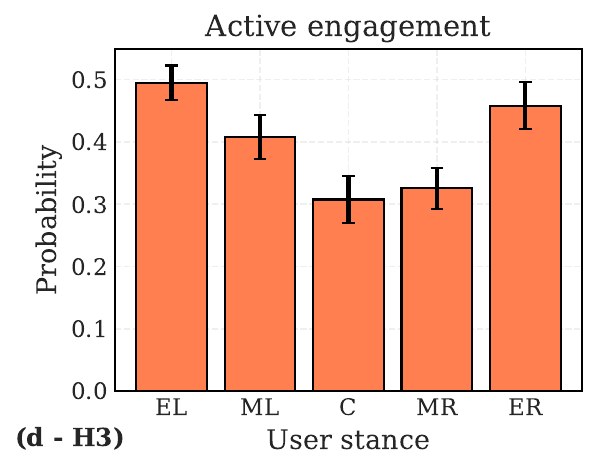}

    

    \caption{Behavioral Parameter Estimation (E1): (a) aggregated user opinion on the topics, (b) Position bias (H1), (c) Stance alignment in clicking (H2), (d) Active engagement (H3). 
    Error bars indicate standard deviations obtained through bootstrap with $1{,}000$ user-level resamples.}
    \label{fig:user_behavior}
\end{figure*}
\subsection{Participant recruitment and News selection}

At the core of our framework is an online platform that measures user interactions with ranked news content. 
We implement the platform using oTree and recruit participants with Prolific.
While simpler than similar platforms~\cite{combs2023reducing,difranzo2018upstanding,jagayatuser,butler2024mis,theophilou2024gender}, it allows for the estimation of the quantities defined in Sec.~\ref{sec:model} and testing model predictions under different ranking conditions.
In this section, we first describe the participant recruitment process and news selection strategy. In Sec. \ref{sec:exp_design}, we then describe the interaction flow in the platform and the design of our experiments.

\subsubsection{Participant Recruitment}
Participants were recruited on Prolific, selecting only US residents with an approval rate above 90\%. Using the ``quota sample'' setting, we selected a balanced number of participants by gender and US political affiliation. We did not control for age as all platforms have a different age prevalence; our data showed in our case the distribution of age was centered around 40. 
Each participant had up to 40 minutes to complete their tasks; during this time, they were shown the terms and conditions, reported their political stance, and interacted with ranked lists of news items. They received £1.50 for completing the task ( $\sim$£9 per hour, based on average completion time).

\subsubsection{News Selection}\label{sec:news_selection}
For each of the four tasks, we selected two news items per political stance ($N=10$ in total) from Ground.news~\footnote{\url{https://ground.news/}}; the website groups news about the same event from multiple outlets and categorizes these outlets into seven political stance categories, with clear stance labels (Far Left, Left, Lean Left, Center, Lean Right, Right, and Far Right).

The political stance of news item $n$, $s_n$, was obtained from the Ground.news categorization of its source. Since we use five stances $\{-2,-1,0,1,2\}$ rather than the seven of Ground.news, we mapped them as follows: $-2$ for far left/left, $-1$ for leaning left, $0$ for center, $1$ for leaning right, and $2$ for right/far right.

To maximize experimental control, for each user interaction, news items shown to participants were all about the \emph{same specific event}, covered by outlets spanning the political spectrum $S$. 


\subsection{Experimental Design}\label{sec:exp_design}

\subsubsection{Participant Interaction Flow}
The core of the experiments comprised four tasks. In each task, participants reported their stance on a specific topic and interacted with related news items. The topics were ``gender policies'', ``vaccination'', ``immigration'', and ``climate change'', selected as controversial (hard news) topics~\cite{bakshy2015exposure}. Task order was randomized for each participant to mitigate order effects; each task involves three sequential steps, with one screen per step
(see Appendix~\ref{app:Screenshots}):\\
(1) \emph{User Stance}: 
Participants read the description of one of the four topics and a description of the typical left/center/right stance on that topic. They then reported their own stance on a 5-item scale (Strongly Left, Leaning Left, Center, Leaning Right, Strongly Right), which we mapped onto the -2 to +2 scale of $S$. \\
(2) \emph{Ranked List and Clicking}: Participants were shown a scrollable ranked list of $N=10$ news items about a topic-related event. Each item displays a title and a left/center/right label. Participants clicked on exactly one item. See Fig.~\ref{appendix:fig:ss_2} for an example. To account for behavioral nudges induced by the interface, we experimentally tested minor variations for interface design; in particular, we tested showing the source itself rather than the label, and smaller/bigger boxes for the news. None of these variations has produced a significant effect on the measured user behavior.\\
(3) \emph{Highlighting (Active Engagement)}: The full article was displayed together with the name of the source outlet. Participants initially saw the first $\sim 1,000$ characters and could click on a ``Read More'' button to reveal the rest of the article. On the same screen, they reported whether they would \emph{like}, \emph{share}, \emph{like and share}, or \emph{do nothing} if reading this article on social media. 

\subsubsection{Behavioral Parameter Estimation (E1)}
\label{sec:E1}
The goal of this experiment is to validate hypotheses H1--H3 (Sec.~\ref{sec:user_behavior}) and to estimate key model parameters for the simulation (Fig.~\ref{fig:overview}).

In this experiment, we collected interactions from 432 participants (May–June 2025) who completed all tasks described in Sec.~\ref{sec:exp_design} using a ranked list that was not updated based on user activity. Instead, a 
randomized ranking was generated for each user and task, with each participant interacting once per topic. Also, the topic order was randomly selected for each participant. 

We used data collected in this condition to obtain max-likelihood estimates of user-stance distributions $\hat{D}_{s_u}$, position bias $\hat{\beta}$, stance-conditioned click $\hat{C}_{s_n,s_u}$, and highlight probabilities $\hat{H}_{s_n,s_u}$. 


\subsubsection{Feedback Loop Experiment (E2)}\label{sec:E2}

Our second experiment involved human participants using dynamically updated rankings. For each topic--configuration pair, we conducted three independent ranking trajectories. Each trajectory was initialized at random and updated independently as participants interacted with it. The design therefore consists of $24$ trajectories in total: $2$ configurations $\times$ $4$ topics $\times$ $3$ trajectories per topic--configuration pair.

For each topic-specific task, each participant was randomly assigned to one of the two algorithmic configurations and to one of the three trajectories for that topic--configuration pair. Participants were shown the ranked list without information about the algorithm that generated it. As in E1, the list was presented simply as ``results of a search'' (Fig.~\ref{appendix:fig:ss_2}), making the conditions comparable.

\begin{figure}[t!]
    \centering
    \includegraphics[width=0.48\columnwidth]{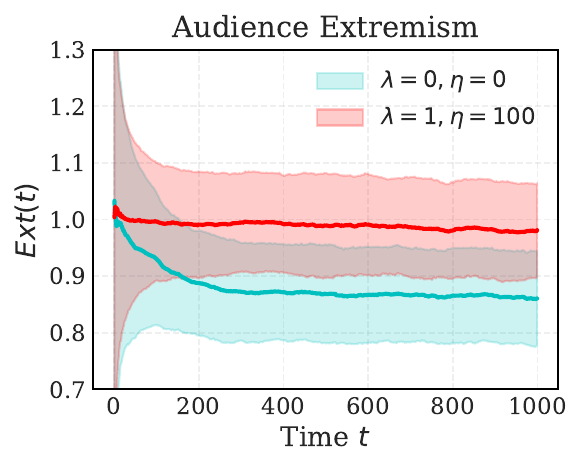}
    \includegraphics[width=0.48\columnwidth]{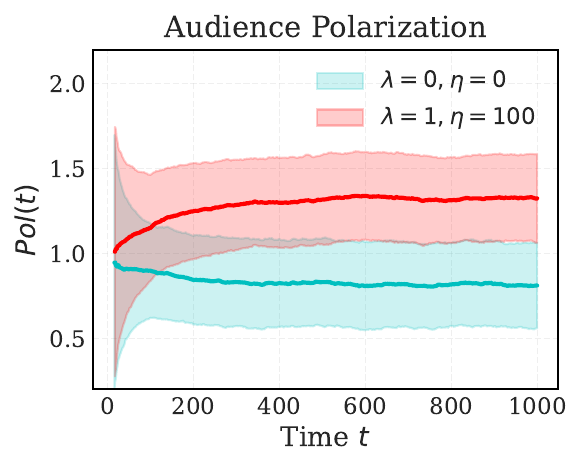}
    \caption{Simulated Audience Extremism (left) and Audience Polarization (right) time series, calculated on a window of $|W|=200$ clicks over $1{,}000$ independent simulations for two algorithmic configurations: $\lambda=0,\eta=0$ (blue) and $\lambda=1,\eta=100$ (red), using overall user behavior. Solid lines indicate means and shaded areas indicate standard deviations.}
    \label{fig:time_metrics}
\end{figure}

To ensure sufficient interactions for the metrics to stabilize, given the convergence pattern observed in the simulations (Fig.~\ref{fig:time_metrics}), we recruited a total of $1{,}534$ participants, yielding approximately $253$ interactions per trajectory.




\label{app:evolution}

\begin{figure}[t!]
    \centering
    \includegraphics[width=0.495\columnwidth]{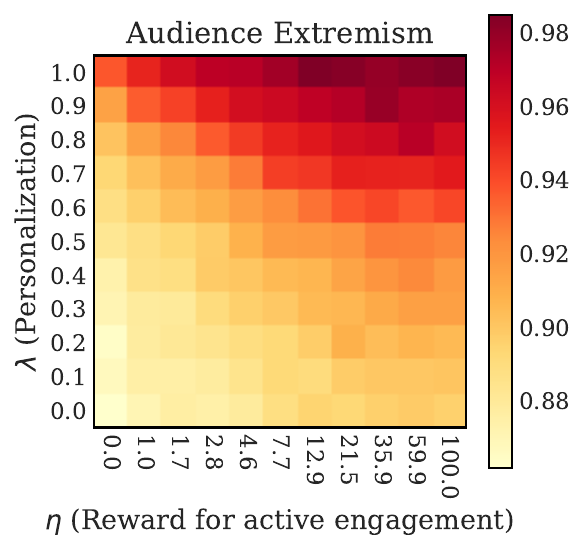}
    \includegraphics[width=0.495\columnwidth]{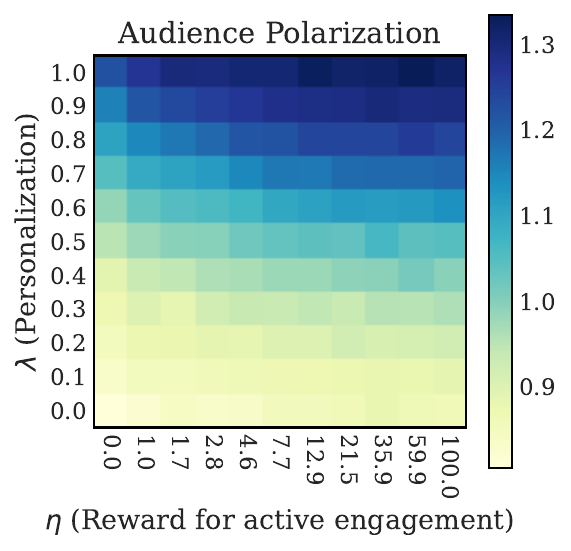}
\caption{Simulated values of Audience Extremism and Audience Polarization for varying personalization ($\lambda$, y axis, linear scale) and active engagement reward ($\eta$, x axis, logarithmic scale), using overall user behavior. Cells show the average of Eq.~\ref{eq:ext} and Eq.~\ref{eq:pol} over $1{,}000$ independent simulations.}
    \label{fig:simulation_grid}
\end{figure}

\section{Results}\label{sec:results}

\begin{figure*}[ht!]
    \centering
    \includegraphics[width=0.9\textwidth]{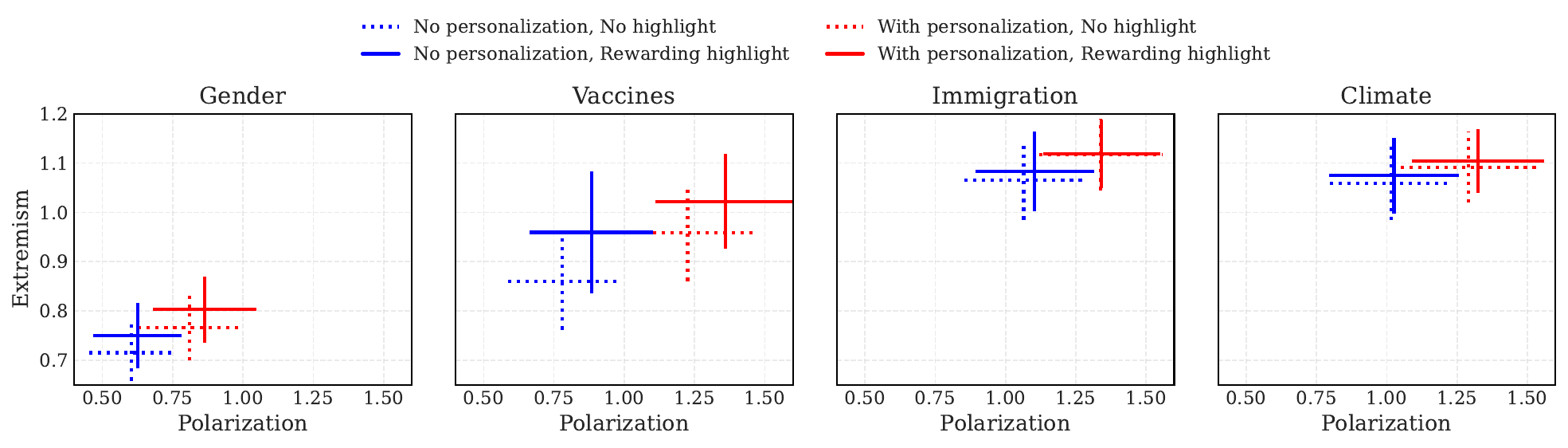}
    \caption{Topic-level simulations of Audience Extremism and Audience Polarization for the four corner algorithmic configurations~(Sec.~\ref{sec:simulation}). Crosses indicate mean $\pm$ standard deviation over $1{,}000$ independent simulations.}
    \label{fig:topic_simulation_corners}
\end{figure*}

\subsection{Behavioral Parameter Estimation (E1)}\label{sec:res_E1}

Fig.~\ref{fig:user_behavior} shows the main results.
The user stance distribution $\hat{D}_{s_u}$ across all topics is slightly left-skewed and concentrated toward the extremes of the political spectrum (Fig.~\ref{fig:user_behavior}(a)).
See App.~\ref{app:GeneralStance} and \ref{app:CandH_tendencies} for a detailed analysis of topic-specific stance comparisons.
    
\emph{Position bias (H1).}
Fig.~\ref{fig:user_behavior}(b) shows the empirical click probability as a function of news rank, with an average position-bias parameter $\hat{\beta}=1.09 \pm 0.01$. This confirms a mild but consistent position bias: higher-ranked items are more likely to be clicked.

\emph{Stance alignment in clicking (H2).}
Fig.~\ref{fig:user_behavior}(c) displays the lift in same-stance clicking relative to a uniform baseline ($L_{s_u} = \hat{C}_{s_u,s_u}/0.2$). Users of all stances show assortative clicking, with center users exhibiting the strongest alignment ($L_{s_u} \approx 2$), followed by extreme users (about $1.5 \times$ the random baseline). 
These results confirm that users prefer news aligned with their own stance.
This pattern is also observed, and is even stronger, when clicks are coded using participants' perceived stance of the clicked article rather than the source-based stance label (App.~\ref{app:NewsPerception}).

\emph{Active engagement (H3).}
Fig.~\ref{fig:user_behavior}(d) shows the probability of highlight (like/share) after clicking, marginalizing over news stance. We observe that engagement follows a U-shaped pattern: extreme-left users engage most ($\sim50\%$), followed by extreme-right ($\sim47\%$), with lower rates for moderates and center users ($30\text{--}40\%$). 
The U-shaped pattern is statistically significant and consistent with previously observed polarization-related engagement patterns, supporting H3.

In summary, these results support the three hypotheses H1--H3.

\subsection{Simulation Results}\label{sec:results_simulation}
Fig.~\ref{fig:time_metrics} shows the evolution of Audience Extremism $Ext(t)$ and Audience Polarization $Pol(t)$ with the number of interactions, for the two corner cases $\lambda=0,\eta=0$ and $\lambda=1,\eta=100$. As users interact with the ranked list, item popularity increasingly reflects their clicking and highlighting tendencies. 
Both metrics have standard deviations of about $0.1$ across independent simulations, due to the different sampled sequences of user stances and interactions. The average values stabilize after roughly $200$ to $300$ interactions.

Based on these simulations, we used approximately $250$ interactions per trajectory in E2. This choice balances budget constraints with the need to reduce metric variability and reach the stable average behavior observed in simulation.

Fig.~\ref{fig:simulation_grid} shows the value of Audience Extremism and Polarization metrics across a grid of algorithm parameter values. Both metrics increase monotonically with $\lambda$ and $\eta$ within the explored parameter range. This indicates that increasing personalization or highlight reward produces a shift toward more ideologically extreme and polarized news consumption in the simulations.

The effect of personalization is particularly pronounced for polarization, which grows faster along the $\lambda$ axis.  
This is in agreement with the interpretation of $\lambda$ as the degree to which users’ rankings depend on others with similar opinions, effectively reinforcing the stance alignment in clicking (H2).
To examine topic-specific variation while isolating the contributions of $\lambda$ and $\eta$, we examined the four scenarios described in Sec.~\ref{sec:simulation} for each topic separately (see Fig.~\ref{fig:topic_simulation_corners}). The differences between topics reflect differences in the estimated parameters $\hat{\beta}$, $\hat{C}_{s_n,s_u}$, and $\hat{H}_{s_n,s_u}$ obtained from E1 at the topic level. 

Topic-specific outcomes can be interpreted directly from estimated parameters. When $\eta=0$, ranking dynamics are driven by clicks, so cross-topic differences mainly reflect the estimated user stance distribution, position bias, and stance-conditioned clicking tendencies. When $\eta>0$, highlighting determines which clicked items receive additional popularity, making topics with engagement concentrated on extreme content more sensitive to engagement reward.

Despite variability across the $1{,}000$ independent simulations, the condition combining personalization and highlight reward consistently produced the highest Audience Extremism and Audience Polarization across all topics. Furthermore, 
personalization but no highlight reward produced higher polarization than 
no personalization but rewarding highlights.  
It resulted in comparable (Gender, Vaccines) or even higher (Immigration, Climate) levels of extremism.  
This confirmed the aggregate-level result that personalization had a stronger effect than highlight rewards on both metrics.

While these patterns held across all topics, each displayed distinct behavior.  
Immigration and Climate Change started with the highest baseline values for both metrics and were less affected by changes in highlight weight.  
In contrast, Gender and Vaccination began at lower levels of extremism and polarization but strongly responded to variations in $\lambda$ and $\eta$.  
The Vaccination topic, in particular, showed the highest sensitivity to the ranking algorithm—a trend later confirmed in the experiment with human participants.

In summary, these simulation results demonstrate that personalization and engagement-based rewards amplify extremism and polarization in user–algorithm interactions.

\subsection{Feedback Loop Experiment (E2)}\label{sec:res_E2}
In the experiment with dynamically updated rankings, we analyzed the effects of different algorithmic configurations in a setting with human participants. In addition, we compared the experimental results with simulation outcomes to validate our framework.

\begin{figure*}[th!]
    \centering
    \includegraphics[width=.85\textwidth]
    {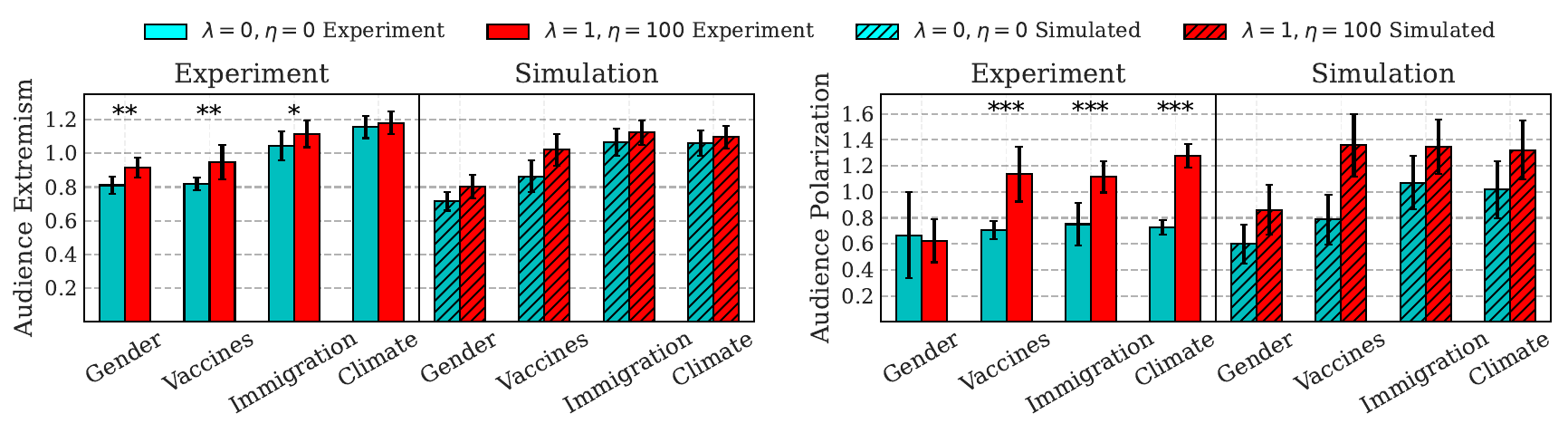}
    \caption{
Audience Extremism (left) and Audience Polarization (right) measured in experiments and simulations under two ranking conditions: no personalization and no highlight reward ($\lambda=0,\eta=0$; blue), and full personalization with strong highlight reward ($\lambda=1,\eta=100$; red). Error bars represent standard deviations across three trajectories for E2 and $1{,}000$ independent simulations. Statistically significant differences between experimental conditions are indicated by *~$p<0.05$, **~$p<0.01$, ***~$p<0.001$, based on one-sided Mann--Whitney U tests using individual clicks. Robustness to block aggregation is discussed in Sec.~\ref{sec:res_E2}.
Both metrics increase under the personalized and engagement-rewarding condition.}
    \label{fig:exp_metrics}
\end{figure*}


\subsubsection{Audience Extremism and Polarization}

Fig.~\ref{fig:exp_metrics} compares Audience Extremism and Audience Polarization under the two conditions described in Sec.~\ref{sec:E2}: a neutral ranking based only on clicks, with no personalization ($\lambda=0,\eta=0$), and a personalized ranking that strongly rewards highlights ($\lambda=1,\eta=100$). 

The metrics were computed in each trajectory over a window of size $w=200$ from the last ranking updates, discarding the first $\sim 50$ interactions to remove transients still dependent on the initial ranking. Averages and standard deviations were then computed across the three trajectories for each configuration and topic.

To test whether the two conditions differ, we used one-sided Mann--Whitney U tests. 
Because clicks within the same trajectory are temporally dependent, tests based on individual clicks may overestimate the effective sample size. We therefore repeated the analysis after aggregating consecutive clicks into blocks of different sizes and computing Eqs.~\ref{eq:ext} and~\ref{eq:pol} within each block. Block size $1$ corresponds to the individual-click test reported in Fig.~\ref{fig:exp_metrics}, while block size $200$ corresponds to aggregation at the trajectory level. With only three trajectories per topic--configuration pair, this level of aggregation does not provide enough samples for statistical significance, but block sizes between $10$ and $50$ produce the same significance pattern as in Fig.~\ref{fig:exp_metrics}.


Audience Extremism increased under the personalized condition with engagement reward for all topics, reaching significance in every topic except Climate Change. Polarization also increased in all topics except Gender.

Across the eight topic--configuration pairs in Fig.~\ref{fig:exp_metrics}, simulations and experiments are strongly correlated: Pearson correlation is $0.88$ ($p=0.004$) for both Audience Extremism and Audience Polarization.
Topics with higher values in the experiment are correctly predicted by the simulation, as are the relative increases across conditions.
 Vaccination exhibits the largest increase in extremism, and Vaccination and Climate Change the largest in polarization. 

\subsubsection{Click Distributions}
The Audience Extremism and Audience Polarization metrics indicate that rewarding active engagement and introducing personalization shifts consumption toward more extreme and like-minded content. To further quantify this effect, we analyzed the click distributions within each user group.


Fig.~\ref{fig:exp_clicks2} shows that, for both left- and right-leaning participants, clicks on same-stance news increased substantially (particularly on extreme items) while clicks on centrist and opposite-stance content decreased. The simulated and experimental click distributions are also strongly correlated ($r=0.97$, $p\sim 10^{-18}$).

For each user group, chi-squared contingency tests confirmed that the click distributions differed significantly across configurations. Relative to the ranking based only on clicks, clicks on extreme same-stance content increased by about $20\%$ for both left- and right-leaning users, largely at the expense of centrist news, which decreased by approximately $15\text{--}20\%$. Center users showed no significant change, consistent with their lower active engagement and the absence of directional bias in personalization.


Evidence of this effect is shown in the popularity-based rankings (Table~\ref{tab:news_ranking}), which report the average position of each news stance across all experimental configurations. When rankings were shared and ``highlighting'' is not rewarded ($\lambda=0, \eta=0$), centrist news were the most popular, appearing near the top of the list (rank~2.4). Under the personalized and engagement-rewarding algorithm ($\lambda=1, \eta=100$), left-leaning users promoted moderate-left and extreme-left content, while right-leaning users elevated extreme-right and moderate-right news. In both cases, centrist and opposing content shifted toward the middle or bottom of the rankings.



\section{Conclusions and Discussion}

\begin{table}[t!]
\centering
\caption{Experiment 2. Average news rank by news stance and configuration. Lower values indicate greater popularity. Darker colors indicate lower average rank. When $\lambda=0$, all user groups see the same ranking.}
\label{tab:news_ranking}
\begin{tabular}{@{}lccccc@{}}
\hline
 & \multicolumn{5}{c}{News stance} \\
\cline{2-6}
Parameters and Partition & EL & ML & C & MR & ER \\
\hline
$\lambda=0, \eta=0$ 
  & \cellcolor{purplebase!28!white}6.0 
  & \cellcolor{purplebase!57!white}4.7 
  & \cellcolor{purplebase!100!white}2.4 
  & \cellcolor{purplebase!22!white}6.7 
  & \cellcolor{purplebase!16!white}7.7 \\

\hline
$\lambda=1, \eta=100$ Left
  & \cellcolor{purplebase!66!white}4.1 
  & \cellcolor{purplebase!82!white}3.3 
  & \cellcolor{purplebase!70!white}3.9 
  & \cellcolor{purplebase!15!white}7.6 
  & \cellcolor{purplebase!10!white}8.7 \\

$\lambda=1, \eta=100$ Center
  & \cellcolor{purplebase!28!white}5.9 
  & \cellcolor{purplebase!27!white}6.1 
  & \cellcolor{purplebase!84!white}3.1 
  & \cellcolor{purplebase!33!white}5.7 
  & \cellcolor{purplebase!24!white}6.6 \\

$\lambda=1, \eta=100$ Right
  & \cellcolor{purplebase!16!white}7.7 
  & \cellcolor{purplebase!22!white}7.0 
  & \cellcolor{purplebase!32!white}5.5 
  & \cellcolor{purplebase!66!white}4.0 
  & \cellcolor{purplebase!84!white}3.2 \\
\hline
\hline

\end{tabular}
\end{table}

\begin{figure}[t!]
    \includegraphics[width=.9\columnwidth]{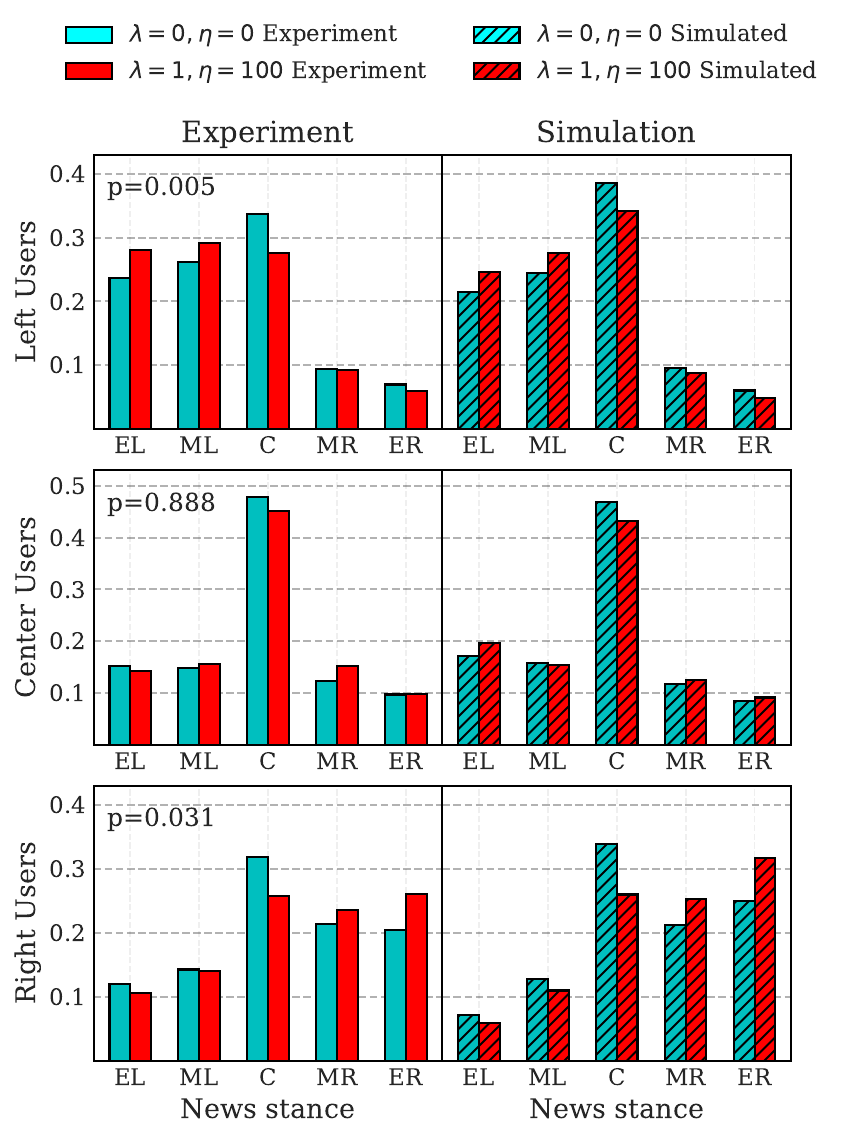}
\caption{Click proportions by user group ($L$, $C$, $R$) and news stance (EL, ML, C, MR, ER) in E2 (left) and simulations (right). Blue and red indicate the two algorithmic configurations. $p$-values correspond to chi-squared contingency tests. Extreme same-stance clicks increased for left- and right-leaning users, mostly at the expense of center news. Experiment--simulation correlation: $r=0.97$ ($p\sim 10^{-18}$).}
\label{fig:exp_clicks2}
\end{figure}

The agreement between E2 and simulations shows that the model in Sec.~\ref{sec:model}, informed by the behavioral estimates from E1, provides accurate predictions of the main experimental patterns ({\textbf{C1}}). Although the simulations are not intended to provide exact quantitative predictions, they reproduce the key effects observed in E2 and correlate strongly with the experimental results (Figs.~\ref{fig:exp_metrics} and~\ref{fig:exp_clicks2}).

The E2 results show that changing from a neutral ranking ($\lambda=0,\eta=0$) to a personalized and engagement-rewarding ranking ($\lambda=1,\eta=100$) produces systematically distinct user behavior, establishing the causal effect of this combined algorithmic configuration relative to the neutral baseline ({\textbf{C2}}). The separate roles of personalization and engagement reward are examined through simulations varying $\lambda$ and $\eta$ systematically. Experimentally, this amplification corresponds to an approximate 20\% increase in clicks toward extreme content from both left- and right-leaning users, and a 13\% increase in clicks to same-stance news among all users.

Combining behavioral estimates, data-driven simulations, and experimental measurements allows us to identify the mechanism behind the amplification of ideologically extreme and polarized consumption ({\textbf{C3}}). E1 confirms three behavioral patterns: (H1) users prefer items ranked near the top of the list, (H2) users tend to click on ideologically aligned content, and (H3) users with more extreme views are more likely to engage actively.

When personalization is introduced, stance-aligned clicks (H2) have a stronger effect on item popularity because they are less counterbalanced by clicks from users with different views. Engagement rewards intensify this effect, since users with more extreme views contribute disproportionately to popularity through active engagement (H3), pushing like-minded extreme content higher in the ranking. Position bias (H1) then reinforces the feedback loop, as content that has moved upward becomes more likely to be clicked in subsequent interactions. As a result, consumption among left- and right-leaning users shifts toward extreme same-stance content, while attention to centrist or opposing content declines.

The simulations further show how the two algorithmic parameters shape these outcomes. Varying $\lambda$ and $\eta$ systematically changes the resulting click patterns, with personalization having the strongest effect on polarization and engagement reward increasing the visibility of ideologically extreme content.

These results suggest that personalization and engagement optimization, while effective at capturing attention, can unintentionally intensify ideologically extreme and polarized consumption. Our model provides a possible mechanistic explanation for engagement trends observed on major platforms such as Facebook. In particular, the 2018 News Feed update under the concept of \emph{Meaningful Social Interactions}~\citep{wsj_2021} increased the weight of active engagement signals such as comments, shares, and reactions, while down-weighting interactions from groups with non-friends and from strangers or small pages. In our framework, these changes can be interpreted as directionally consistent with increasing engagement reward and personalization. Consistent with the analysis of~\citet{fraxanet-EPJ25}, such changes could raise overall interaction levels while also amplifying ideological segregation and the visibility of extreme content.

Our experiments focus on hard-news content, which is inherently controversial and often displays a U-shaped engagement profile across ideology~\citep{bakshy2015exposure}. The model also generates predictions for settings with flatter or qualitatively different engagement profiles. Platforms such as Wikipedia, Reddit, and Bluesky provide useful comparison cases because prior work suggests they involve weaker or different forms of ideological sorting, engagement concentration, or algorithmic amplification~\citep{greenstein2016ideological,waller2021quantifying,bluesky}. These settings could be used to test whether weaker engagement concentration leads to weaker amplification, as predicted by the model.

Future work is needed to examine how alternative ranking criteria might balance personalization with exposure diversity to mitigate these effects. Furthermore, this work highlights the need for transparency from social media platforms, as increasing evidence shows that their algorithms can influence society.

\paragraph{Limitations:}

We acknowledge that our methodological framework has some limitations. First, we cannot be certain that participants behaved as they would on actual social media platforms, despite our efforts to emulate a natural browsing experience. Our framework is designed to isolate a fundamental mechanism and, as a result, has more limited ecological validity than field-based studies~\cite{doi:10.1126/science.adu5584}. This design choice allows for a clear characterization of the amplification mechanism: because users with more extreme views engage more actively, engagement-based ranking increases the visibility of ideologically extreme content to a broader audience, which in turn induces users with less extreme views to interact with such content. Moreover, compared to real-world settings where platforms rely on richer engagement signals, such as comments, our framework restricts users to at most one highlight per interaction, which is considerably more limiting. This suggests that the mechanism identified here could be stronger in less constrained and more realistic environments, although this remains an empirical question for future work.

Second, our experiments focused on conditions where personalization and engagement reward operate jointly, reflecting realistic platform settings. Although our simulations vary $\lambda$ and $\eta$ systematically, disentangling their independent and interactive effects experimentally would require testing all four conditions corresponding to the combinations of personalization and engagement reward.

\paragraph{Ethical Use of Data and Informed Consent:}

This study complies with ACM's Publications Policy on Research Involving Human Participants and Subjects (2021). The authors affirm that all data collection and analysis followed institutional and disciplinary ethical standards. All procedures involving human participants were reviewed and approved by the appropriate institutional ethics board, and informed consent was obtained prior to participation. The consent form stated that ``participating in this study does not entail risks greater than those ordinarily encountered in daily life''. Furthermore, participants had already disclosed their political leanings to Prolific, which was used for quota sampling.

\section*{Acknowledgements}
\begingroup
\emergencystretch=2em

We thank Francesco Sobbrio for useful feedback.
This work is part of the Maria de Maeztu Units of Excellence Programme
CEX2021-\allowbreak001195-M, funded by
MICIU/AEI/\allowbreak10.13039/\allowbreak501100011033.
G. L. received funding from an ICREA Academia grant
PID2022-\allowbreak137908NB-I00 funded by
MICIN/AEI/\allowbreak10.13039/\allowbreak501100011033 and by ERDF/EU
``A way of making Europe'', and the Severo Ochoa Programme for Centres
of Excellence in R\&D (Barcelona School of Economics
CEX2019-\allowbreak000915-S) funded by
MCIN/AEI/\allowbreak10.13039/\allowbreak501100011033.
F. G. acknowledges funding from grant
PID2023-\allowbreak153318NB-I00 funded by MICIU/AEI/UE and from the
Severo Ochoa Programme for Centres of Excellence in R\&D
(Barcelona School of Economics CEX2019-\allowbreak000915-S), funded by
MCIN/AEI/\allowbreak10.13039/\allowbreak501100011033.
A. K. acknowledges support from the Serra Húnter Programme
(Generalitat de Catalunya) as a Serra Húnter Associate Professor.

\par
\endgroup





\pagebreak

\bibliographystyle{ACM-Reference-Format}
\bibliography{sample-base}

\appendix

\renewcommand{\thefigure}{\Alph{section}\arabic{figure}}
\setcounter{figure}{0}

\renewcommand{\thetable}{\Alph{section}\arabic{table}}
\setcounter{table}{0}

\section{User Interface Screenshots}
\label{app:Screenshots}
\begin{figure}[ht]
    \centering
    \includegraphics[width=0.85\columnwidth]{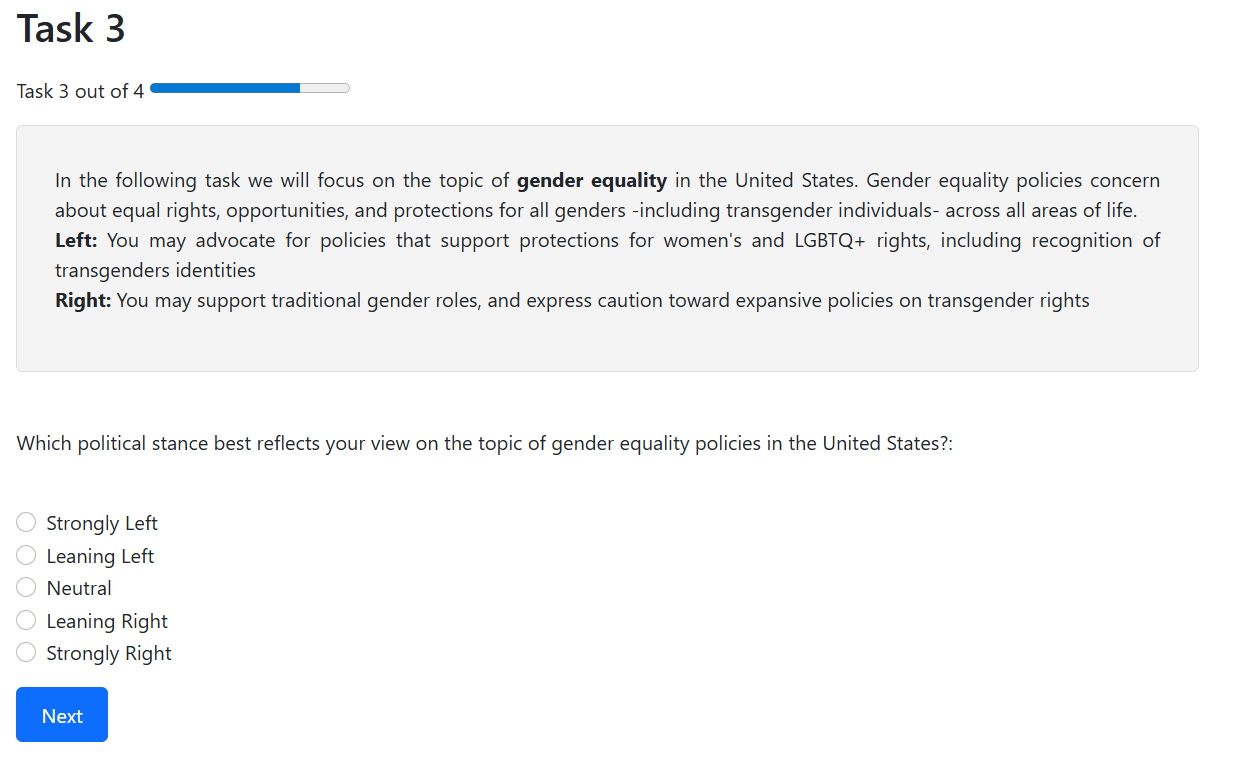}
    \caption{Instance of a page of the "User Positioning" step.}
    \label{appendix:fig:ss_1}
\end{figure}

\begin{figure}[ht]
    \centering
    \includegraphics[width=0.85\columnwidth]{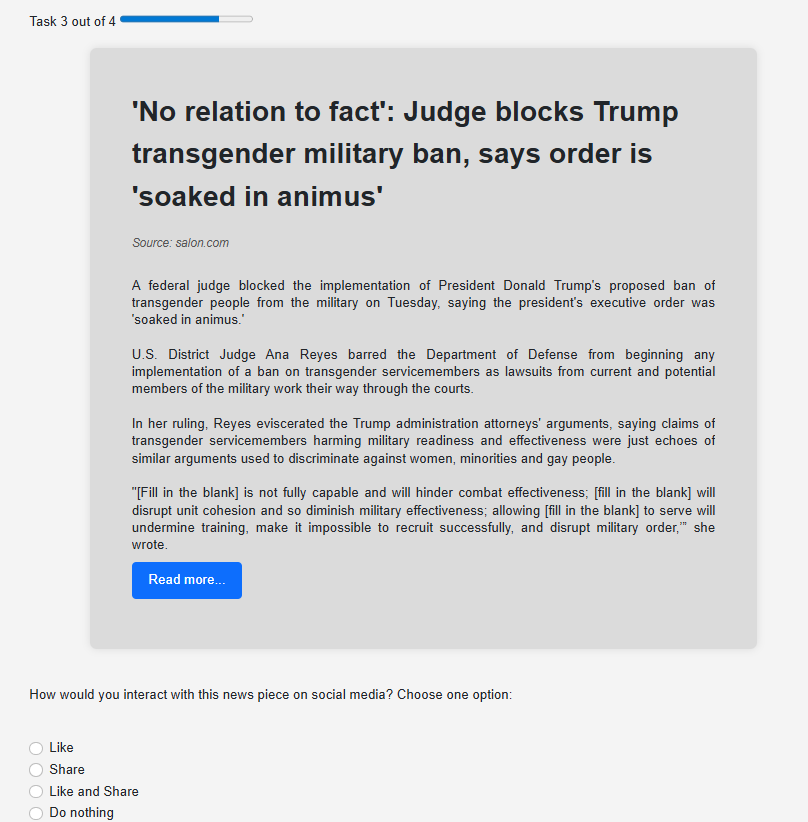}
    \caption{An example of a page that participants see during the "Active Engagement" stage of a task.}    \label{appendix:fig:ss_3}
\end{figure}

In each of the topic-specific tasks, users are shown a screen for: (1) revealing their opinion on the topic (Fig.~\ref{appendix:fig:ss_1}), (2) interacting with the ranked list (Fig. \ref{appendix:fig:ss_2}), then (3) reading the article and deciding how to engage with it (Fig.~\ref{appendix:fig:ss_3}).

\section{User General stance}
\label{app:GeneralStance} 
At the end of the tasks, we assess users' general political stance, which might differ from topic-specific ones. 
We do this in two steps: first asking their stance on the Left, Center, Right axis, then the strength of their opinion, to distinguish moderate from extreme users.

Comparing the results with the topic-specific political stance, we observe a broad agreement between the two, with most of the matrix mass on the diagonal. Nonetheless, we see that quite often users who self-define as ``generally moderate'' also consider themselves extremists about the specific topics. This behavior is observed for both left and right users, while center users' stances on the topic are quite spread out in all possible stances.

This observation shows that it is important to distinguish between the overall stance of a user and their stance on specific cultural or social issues, especially when dealing with divisive topics such as the ones used in the experiments. It also indicates that the relation between extremism and active engagement may not only be driven by extremist users, but also moderate ones that decide to actively engage when they assume a stronger political position on specific issues.

\begin{figure}[ht]
    \centering
    \includegraphics[width=0.45\columnwidth]{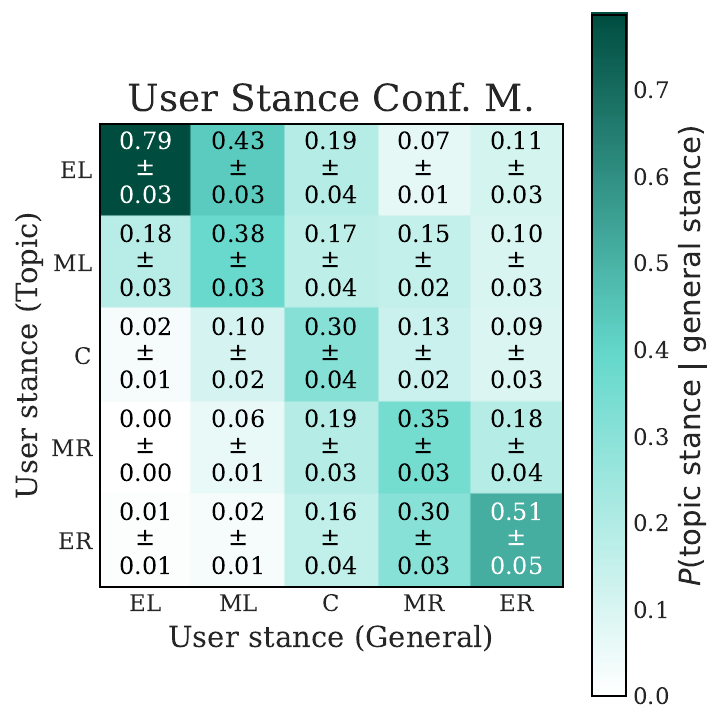}
    \hfill
    \includegraphics[width=0.45\columnwidth]{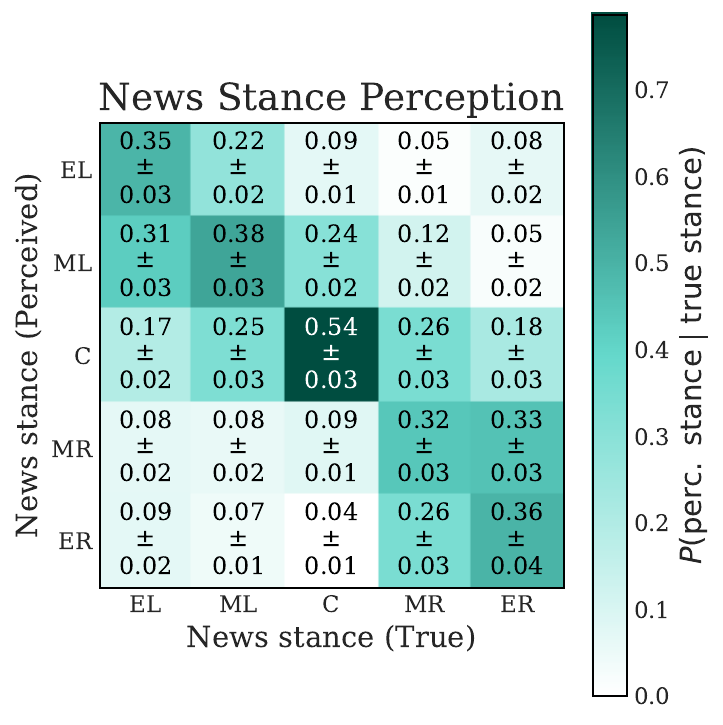}
    \caption{Column-normalized agreement matrices. Left: agreement between users’ general stance and their stance on specific topics. Right: agreement between the true stance of news articles and the stance perceived by users.}
    \label{appendix:fig:stance_agreement}
\end{figure}

\section{News perception}
\label{app:NewsPerception}
 To assess the alignment of the news items with the participants' perception of it, in the final page of the interface (while the clicked article is still being shown), users were asked to position the stance of the article in the political spectrum $S$.
 The confusion matrix of these measures indicates a broad agreement between the labels assigned to news (according to Sec.\ref{sec:news_selection}) and what users perceive. Nonetheless, extreme content seems to be often perceived as moderate, while moderate content is less often perceived as extreme. Combined with the findings of general and topic-specific stances, we can see that the distinction between extreme and moderate stances is a bit blurry, although a general correlation pattern emerges. 
Still, despite this variability, users, on average, agree with the stances assigned to news items, adding robustness to our work. As a robustness check, we repeated the click-matrix analysis using participants' perceived stance instead of source labels. In this case, the same qualitative pattern emerged, with even stronger assortative clicking.



\section{Click and Highlight tendencies in each topic}
\label{app:CandH_tendencies}
\begin{figure}[ht]
    \centering
    \includegraphics[width=\columnwidth]{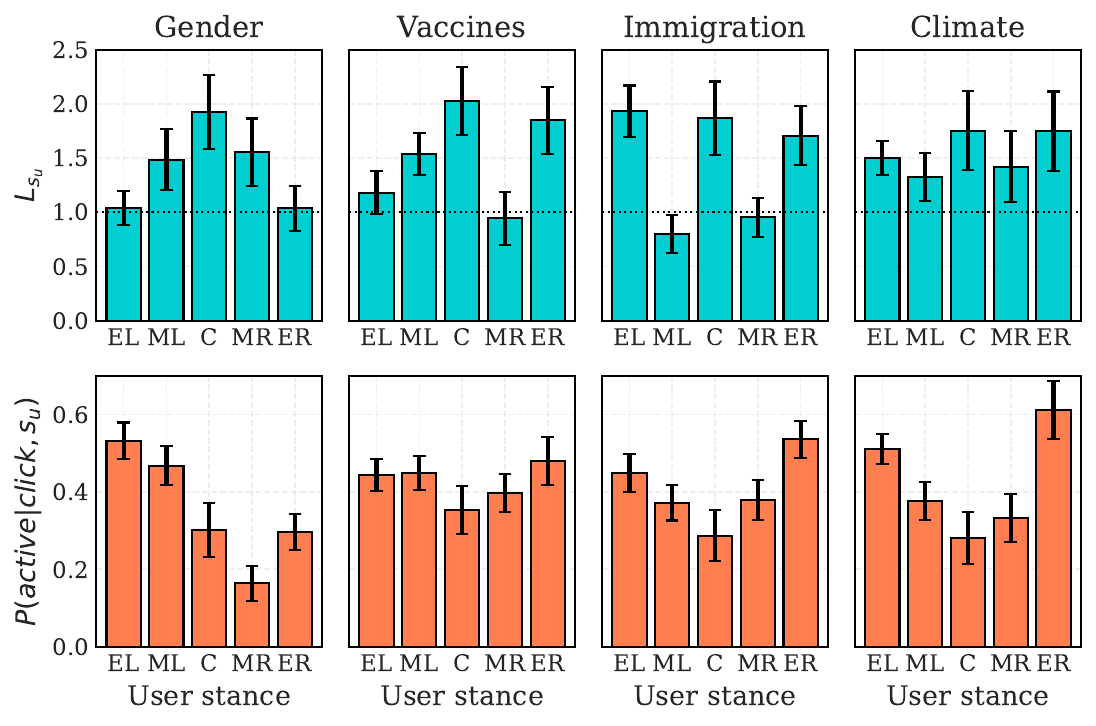}
    \caption{Stance alignment in clicking, and ``highlight'' probabilities measured in each topic independently.}
    \label{appendix:fig:topic_lift_and_vshape}
    
\end{figure}

The top panels of Fig.~\ref{appendix:fig:topic_lift_and_vshape} show the lift in same-stance clicking relative to a uniform baseline ($L_{s_u} = \hat{C}_{s_u,s_u}/0.2$) per topic (in analogy to Fig.~\ref{fig:user_behavior}c. We observe that the same-stance-clicking posited by H2 is broadly
present for all topics, although with topic-specific variations: for the topic gender, for instance, extreme users do not tend to click on extreme news as much as in other topics.\\
Similar observations can be drawn for the U-shape posited by H3, which are also observed for each topic individually (bottom panels in Fig.~\ref{appendix:fig:topic_lift_and_vshape} in analogy to Fig.~\ref{fig:user_behavior}b). The only exception is again for Gender, where right-leaning users seem to be less active in general.\\
We observe that all these distributions are significancy different (based on a random comparison with 1000 samples) from a random baseline. 
This indicates that our hypotheses are valid in most cases. However, it is crucial to account for topic-specific changes to have a more accurate view of user behavior.






\section{Estimated Parameters}
\begin{figure}[ht]
    \centering
    \includegraphics[width=1.\columnwidth]{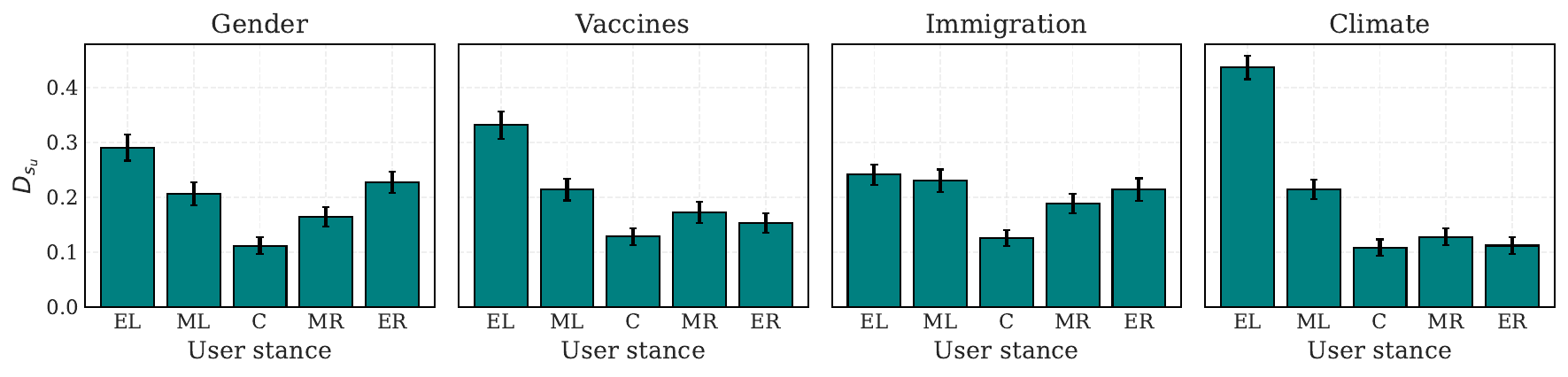}
    \caption{Distribution $\hat{D}_{s_u}$ of user stances on all topics}
    \label{appendix:fig:topic_user_stances}
\end{figure}

\begin{figure}[ht]
    \centering
    \includegraphics[width=1.\columnwidth]{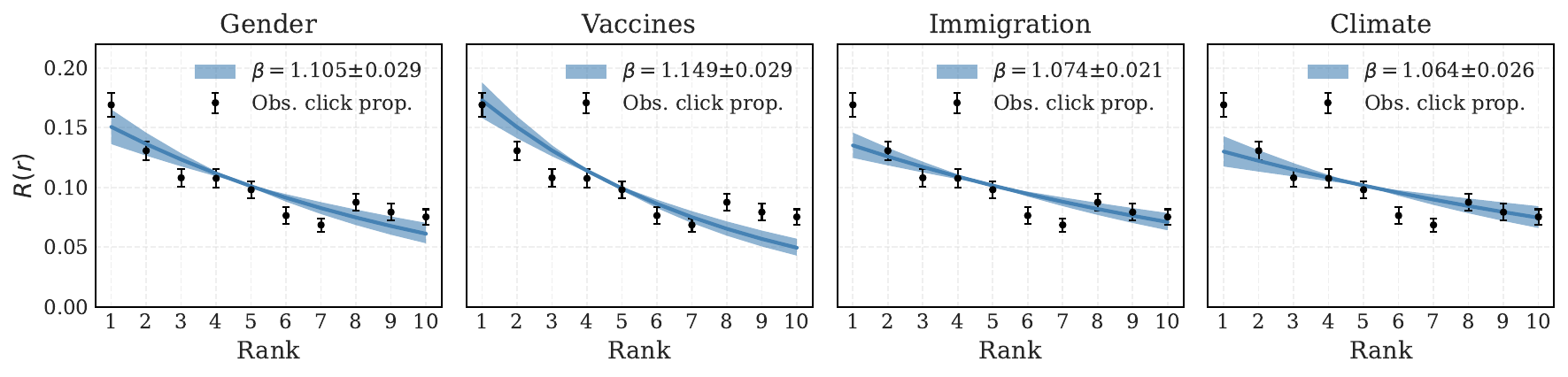}
    \caption{Position bias $\hat{\beta}$ measured in each topic independently}
    \label{appendix:fig:topic_beta}
\end{figure}

\begin{figure}[ht]
    \centering
    \includegraphics[width=1.\columnwidth]{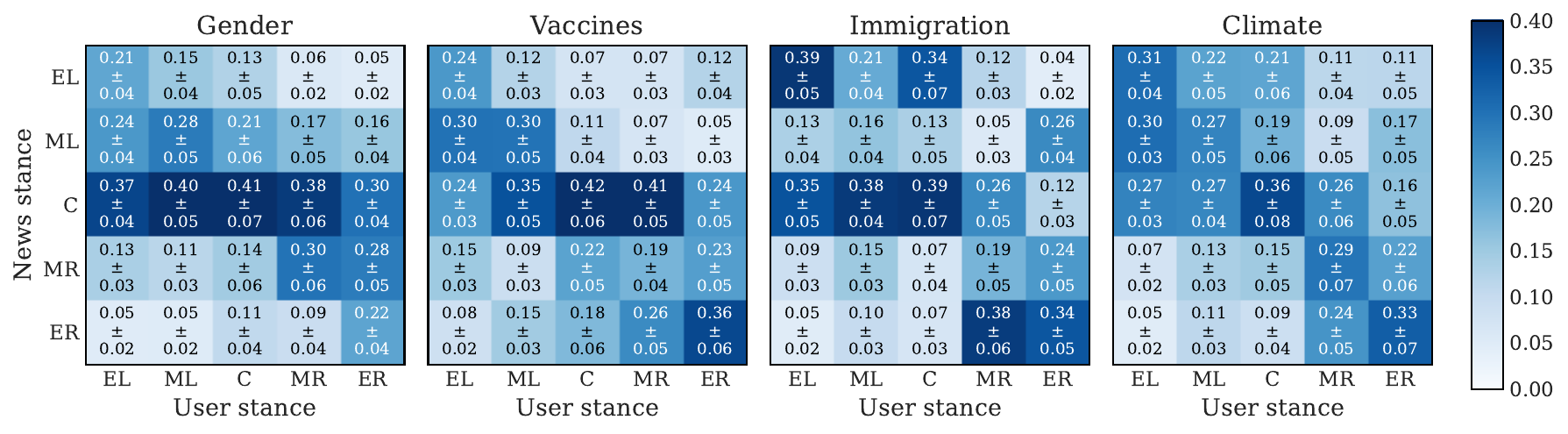}
    \caption{Click matrices $\hat{C}_{s_n,s_u}$ for all topics}
    \label{appendix:fig:topic_cm}
\end{figure}

\begin{figure}[ht]
    \centering
    \includegraphics[width=1.\columnwidth]{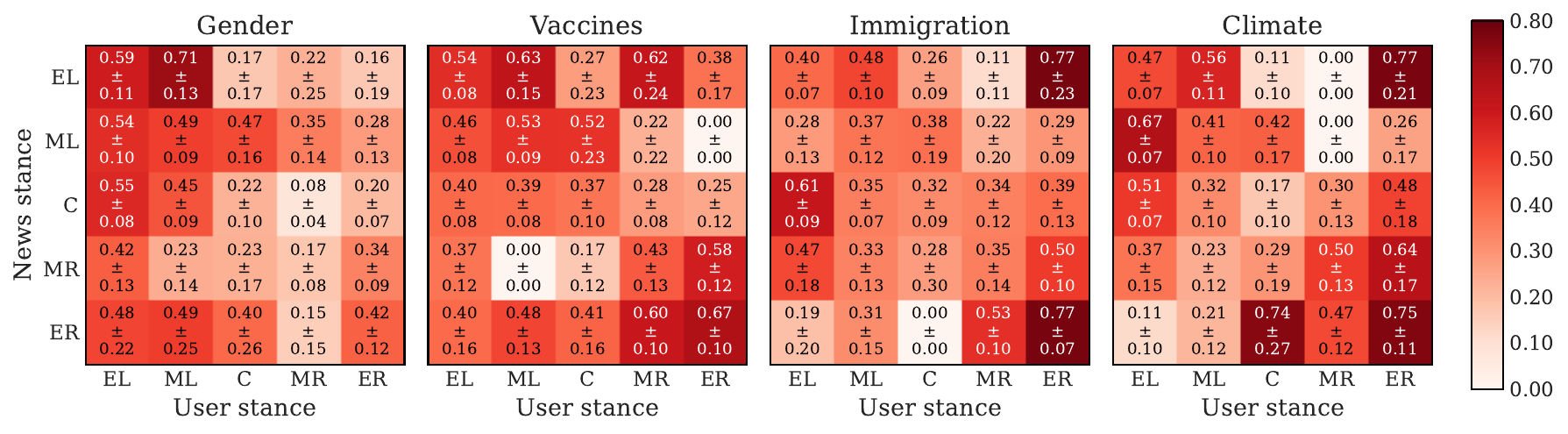}
    \caption{Highlight matrices $\hat{H}_{s_n,s_u}$ for all topics}
    \label{appendix:fig:topic_hm}
\end{figure}

In this section, we show the input values per topic extracted from the static ranking experiment, which were used as input for the simulations. In particular,  
Fig.~\ref{appendix:fig:topic_user_stances}, gives the distribution of the user stances 
$\hat{D}_{s_u}$, The position bias $\hat{\beta}$ is shown in Fig.~\ref{appendix:fig:topic_beta}, the stance-conditioned click probabilities $\hat{C}_{s_n,s_u}$ in Fig.~\ref{appendix:fig:topic_cm}, and finaly the post-click highlight probabilities $\hat{H}_{s_n,s_u}$ in 
Fig.~\ref{appendix:fig:topic_hm}.

\end{document}